\documentclass[aps,amsfonts,prl,twocolumn,showpacs]{revtex4-1}

\usepackage{subfigure}
\usepackage{wrapfig}
\usepackage{textcomp}
\usepackage{graphicx}
\usepackage{amssymb}
\usepackage{amsmath}
\usepackage{bm}
\usepackage{amsmath, amsthm, amssymb}

\def\beq{\begin{equation}}
\def\eeq{\end{equation}}
\def\bea{\begin{eqnarray}}
\def\eea{\end{eqnarray}}

\begin{document}

\title{Mobile magnetic impurities in a Fermi superfluid: a route to designer molecules}

\author{Sarang Gopalakrishnan$^1$, Colin V. Parker$^2$, and Eugene Demler$^1$}
\affiliation{$^1$Department of Physics, Harvard University, Cambridge MA 02138, USA \\ $^2$James Franck Institute and Department of Physics, University of Chicago, Chicago IL 60637, USA}

\begin{abstract}
A magnetic impurity in a fermionic superfluid hosts bound quasiparticle states known as Yu-Shiba-Rusinov (YSR) states. We argue here that, if the impurity is mobile (i.e., has a finite mass), the impurity and its bound YSR quasiparticle move together as a midgap molecule, which has an unusual ``Mexican-hat'' dispersion that is tunable via the fermion density. We map out the impurity dispersion, which consists of an ``atomic'' branch (in which the impurity is dressed by quasiparticle pairs) and a ``molecular'' branch (in which the impurity binds a quasiparticle). We discuss the experimental realization and detection of midgap Shiba molecules, focusing on Li-Cs mixtures, and comment on the prospects they offer for realizing exotic many-body states. 
\end{abstract}

\maketitle

A key project in ultracold atomic physics~\cite{zwerger_RMP} involves using the richness of atomic structure to create ``designer'' many-body systems---e.g., spins with $SU(N)$ symmetry~\cite{gorshkov_SUN} or bosons in gauge fields~\cite{gaugefield_RMP}---that have no solid-state equivalent. 
Condensed matter physics, meanwhile, has developed the converse project of exploiting many-body correlations to generate quasiparticles (e.g., anyons~\cite{FQHEnobel, *nayak_RMP}) that are qualitatively unlike electrons or atoms. 
Such quasiparticles are usually \emph{excitations}, but might exist, even at zero temperature, at impurities, topological defects, or edges~\cite{fu_kane_vortex, *lutchyn_wires}. 
In real materials, impurities, edges etc. are immobile on the timescales of interest. But ultracold atomic systems do not have this restriction, and in these systems impurities are naturally \emph{mobile}: hence the impurity and its captured quasiparticle can form a coherently moving \emph{molecule}. 
 Binding exotic quasiparticles to mobile impurities offers a new method for designing particles whose dispersion and exchange statistics are inherited from an underlying correlated many-body state. Such ``designer molecules'' can access regimes of few- and many-body physics that are inaccessible by purely atomic or solid-state approaches.

\begin{figure}[b]
\begin{center}
\includegraphics{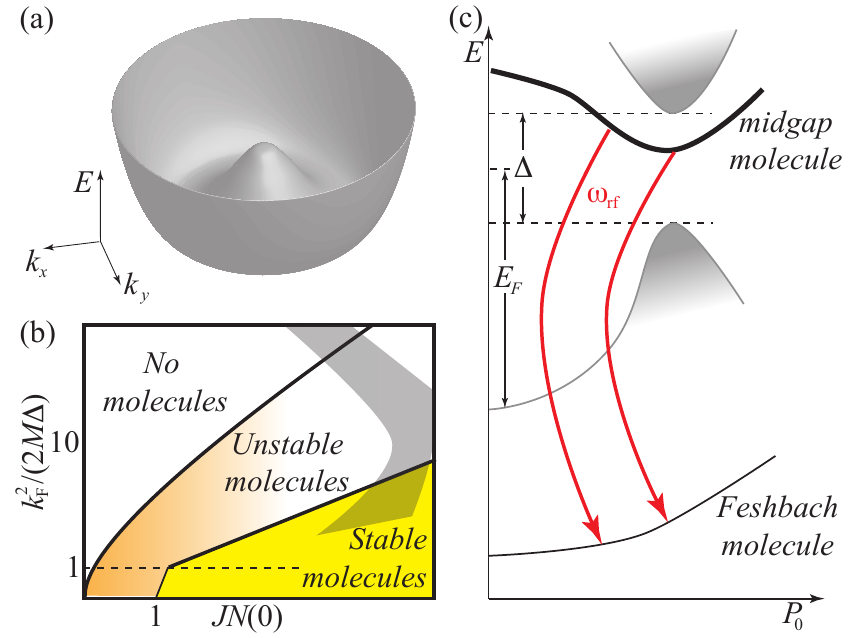}
\caption{(a)~Mexican-hat dispersion of midgap Shiba molecule. (b)~Phase diagram in three dimensions, as a function of impurity mass $M$ and impurity-fermion coupling $J$. As $J$ is increased, the system goes from a phase in which the midgap Shiba molecule does not exist, to one in which it exists as an excited (i.e., unstable) state, and finally one in which it is the ground state. Boundaries are given by Eqs.~\eqref{threshold},~\eqref{transit} (including mass renormalizations as discussed in  text). For heavy impurities (below dashed gray line) one recovers fixed-impurity behavior. Shaded region indicates the achievable parameter regime for Li-Cs mixtures near the 843 G and 880 G heteronuclear resonances~\cite{repp_heteronuclear, colin-pra}. (c)~Dispersion relation, showing midgap Shiba molecule (thick black line), ``under-sea'' Feshbach molecule (thin black line), and Bogoliubov quasiparticles (gray lines). The midgap Shiba molecule's dispersion can be mapped out by driving radio-frequency (rf) transitions from it to the Feshbach molecule.}
\label{summaryfig}
\end{center}
\end{figure}

Here, we consider perhaps the simplest such system, comprising a mobile magnetic impurity in a fermionic superfluid; natural experimental realizations include two-species mixtures (e.g., Li-Cs mixtures) in which one species is fermionic. When the impurity is spatially localized, it binds a midgap quasiparticle state, called a Yu-Shiba-Rusinov (YSR) state~\cite{yu, shiba, rusinov1, salkola, yazdani97, balatsky_RMP, vernier, huihu, sau2013, pientka-glazman, yazdani2013, klinovaja, franz2013}. Depending on the impurity-fermion coupling, the YSR state is either occupied or empty at zero temperature. 
We argue that when the impurity is mobile, it moves together with its quasiparticle state, forming a midgap ``Shiba'' molecule; at strong coupling, this molecule is the ground state of the system (Fig.~\ref{summaryfig}). The midgap Shiba molecule differs from the molecule formed by an impurity in a one-component Fermi gas~\cite{zwierlein-polaron, chevy2006, chevy-lobo, prokofiev, combescot2008}; the midgap Shiba molecule exists deep in the BCS limit, where there are no two-body bound states. Furthermore, the midgap Shiba molecule has an unusual dispersion with a spherical minimum inherited from the Fermi surface (Fig.~\ref{summaryfig}). Spherical dispersion minima have attracted interest in the context of light-induced Rashba spin-orbit coupling, because their high degeneracy enhances interactions, stabilizing exotic correlated phases~\cite{stanescu, *zhai, *mondragon, zhai2, *sg-lamacraft, *ozawa, *barnett, *erez-rudner, *sedrakyan, wilson:meron, *sg:quasi, nigelcooper}. Optically realizing an isotropic Rashba dispersion is challenging~\cite{campbell, *spielman-review}, whereas the dispersion of midgap Shiba molecules in an isotropic system is automatically isotropic.

\begin{figure}[t]
\begin{center}
\includegraphics{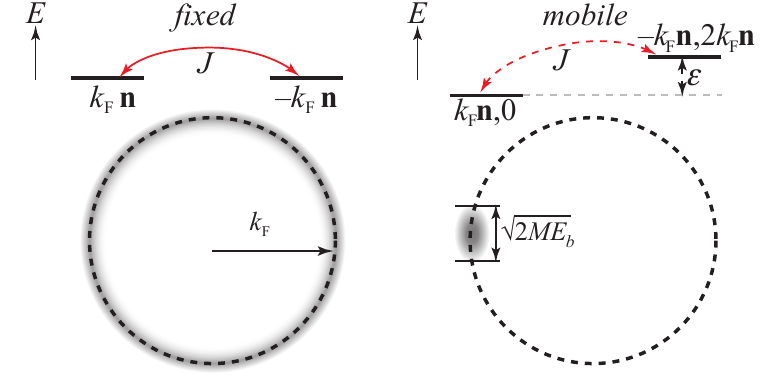}
\caption{Differences between bound states for a fixed impurity (left) and a mobile impurity (right). Scattering across the entire Fermi surface is resonant for the fixed impurity but off-resonant for the mobile impurity owing to recoil (top panel). Consequently, the Shiba state has contributions from all over the Fermi surface for a fixed impurity but only a small patch of the Fermi surface for a mobile impurity (lower panel).} 
\label{structure}
\end{center}
\end{figure}

The parameter controlling the midgap Shiba molecule's unusual properties is the impurity recoil energy, $\mathcal{E} \equiv 2 k_F^2/M$, where $M$ is the impurity mass and $k_F$ is the Fermi momentum~\footnote{Note that $\mathcal{E}$ is also the scale governing the physics of the so-called Kondo polaron~\cite{lamacraft-kondo, *vojta-kondo}.}. For heavy impurities, $\mathcal{E}$ is small compared with the impurity-fermion coupling; therefore, impurity scattering mixes all the states near $k_F$, and the bound-state properties resemble those of a fixed impurity (Fig.~\ref{structure}). However, when $\mathcal{E}$ is \emph{large}, processes scattering a quasiparticle across $2k_F$ are off resonance by $\mathcal{E}$~\cite{lamacraft-kondo, *vojta-kondo}; therefore, the lowest-energy molecular-branch states consist of an impurity with momentum $\sim 0$ and a quasiparticle with momentum $\sim k_F \mathbf{\hat{n}}$ along some \emph{specific} direction $\mathbf{\hat{n}}$. Consequently, when a molecule exists, it must have center-of-mass momentum $k_F$. Since $\mathbf{\hat{n}}$ is arbitrary, the molecular branch has a circular or spherical dispersion minimum by symmetry.

Below, we address the central questions concerning these unusual molecules and polarons. First, we identify critical couplings for the midgap Shiba molecule to exist as (a) an excited state, and (b) the ground state. Second, we compute the effective-mass corrections for both the impurity itself (the ``polaron'') and the midgap Shiba molecule, thus mapping out the full dispersion of the one-impurity problem. Finally, we discuss the regime of validity of our analysis, and propose an experimental method for probing the midgap Shiba molecule. 

\emph{Model}. We consider a system governed by the Hamiltonian 

\bea
\mathcal{H} & = &  \mathbf{P}^2/(2M) + H_{BCS}  +  H_{int}, \\
H_{BCS} & = & \sum\nolimits_{\mathbf{k}} \left[ \sum\nolimits_\sigma \epsilon_k c^\dagger_{\mathbf{k}\sigma} c_{\mathbf{k}\sigma} + \Delta (c^\dagger_{\mathbf{k}\uparrow} c^\dagger_{-\mathbf{k}\downarrow} + \mathrm{h.c.})\right], \nonumber \\
H_{int} & = & \mathcal{V}^{-1} \sum\nolimits_{\mathbf{k k}' \sigma} (V + J \sigma) e^{i \mathbf{(k - k') \cdot X}} (c^\dagger_{\mathbf{k} \sigma} c_{\mathbf{k}' \sigma} + \mathrm{h.c.}) \nonumber
\eea
Here, $\mathbf{X}, \mathbf{P}$ are the impurity position and momentum; $H_{BCS}$ and $H_{int.}$ are respectively the fermionic BCS Hamiltonian and the fermion-impurity interaction; $c_{\mathbf{k}\sigma}$ annihilates a microscopic fermion of momentum $\mathbf{k}$ and spin $\sigma = \pm 1$; $\epsilon_k = v_F (k - k_F)$ is the linearized free fermion dispersion; $\Delta$ is the superconducting gap; $V$ and $J$ are spin-independent and spin-dependent parts of the impurity-fermion interaction; and $\mathcal{V}$ is the system volume. In terms of Bogoliubov quasiparticle operators $\gamma_{\mathbf{k}\uparrow} \equiv u_{\mathbf{k}\uparrow} c_{\mathbf{k}\uparrow} + v_{\mathbf{k}\uparrow} c^\dagger_{-\mathbf{k}\downarrow}$, one can rewrite $H_{BCS} = \sum\nolimits_{\mathbf{k}\sigma} E_k \gamma^\dagger_{\mathbf{k} \sigma} \gamma_{\mathbf{k}\sigma}$; the quasiparticle dispersion is $E_k = \sqrt{\Delta^2 + \epsilon_k^2}$. Conserved quantities under $\mathcal{H}$ are (i)~the total impurity plus fermion momentum, $\mathbf{P}_0$; (ii)~the fermion parity; (iii)~the number difference between $\uparrow$ and $\downarrow$ fermions. We focus on the experimentally relevant three-dimensional case; the one-dimensional case is discussed in the Supplemental Material. 

\emph{Molecular threshold}. We first estimate the threshold for a molecular state to exist, using perturbation theory in $H_{int}$. We express $H_{int}$ in terms of quasiparticles; for simplicity we take $V = 0$:

\bea\label{pert}
H_{int} & = & \mathcal{V}^{-1} \sum\nolimits_{\mathbf{k k}' \sigma} \!\! e^{i \mathbf{(k - k') \cdot X}} \!\left[J \sigma (u_\mathbf{k} u^*_{\mathbf{k}'} + v_\mathbf{k}v^*_{\mathbf{k}'}) \right] \gamma^\dagger_{\mathbf{k}\sigma} \gamma_{\mathbf{k'}\sigma}  \nonumber \\
& & + \sum\nolimits_{\mathbf{k k}' }  e^{i \mathbf{(k - k') \cdot X}} \left[ J \sigma u_\mathbf{k} v_{\mathbf{k}'} \gamma_{\mathbf{k}\downarrow} \gamma_{-\mathbf{k}'\uparrow} \right] + \mathrm{h.c.}
\eea
%
Perturbatively, we are only concerned with states for which $E_\mathbf{k} \simeq \Delta$; for these, $u_\mathbf{k} \approx v_\mathbf{k} \approx 1/\sqrt{2}$
~\footnote{For pure potential scattering, the relevant coherence factor in the first line would be $u_\mathbf{k} u^*_{\mathbf{k}'} - v_\mathbf{k}v^*_{\mathbf{k}'}$, which vanishes for states near the gap, as expected from Anderson's theorem.}. 
Here the $\gamma^\dagger\gamma$ terms involve scattering between the impurity and a quasiparticle; $\gamma^\dagger\gamma^\dagger$ ($\gamma \gamma$) terms create (destroy) quasiparticle pairs.  
Pair creation/destruction inevitably changes the energy by $\sim 2\Delta$, and is off-resonant, whereas the energy change associated with scattering a quasiparticle from one state to another can be arbitrarily small. Therefore, to leading order, we neglect the second line of Eq.~\eqref{pert}; under this approximation the total quasiparticle number is conserved. We wish to look for a bound state of the impurity and one quasiparticle; evidently, this is a two-particle scattering problem with a contact interaction. 
The unusual feature is the ``Mexican-hat'' quasiparticle dispersion: the lowest-energy states with one quasiparticle are those in which the quasiparticle has momentum $\sim k_F \mathbf{\hat{n}}$, where $\mathbf{\hat{n}}$ is an arbitrary unit vector, and the impurity has momentum $\mathbf{P} = 0$; thus, $\mathbf{P}_0 = k_F\mathbf{\hat{n}}$. 
The perturbation couples such a state to other states with impurity momentum $\mathbf{p}$ and quasiparticle momentum $k_F \mathbf{\hat{n}} - \mathbf{p}$. Because states with $|\mathbf{p}| \simeq k_F$ are suppressed by large recoil energy denominators $\sim \mathcal{E}$, we assume $|\mathbf{p}| \ll k_F$. Taking $\mathbf{\hat{n}} = \mathbf{\hat{x}}$, the energy of a state with a given $\mathbf{p}$ is

\beq
\Xi_{\mathbf{p}} \simeq \frac{(p^2)}{2M} + \Delta + \frac{v_F^2}{2 \Delta} p_x^2 = \Delta + \frac{p_x^2}{\left(\frac{1}{2M} + \frac{v_F^2}{2\Delta}\right)^{-1}} + \frac{p_\perp^2}{2M}
\eeq
where $p_\perp \equiv (p_y, p_z)$. This is simply the dispersion of a free particle with an anisotropic mass, $M_y = M_z = M, M_x = (1/M + v_F^2/\Delta)^{-1}$. Thus, the criterion for the midgap Shiba molecule to exist is the same as that for a particle with anisotropic mass, subject to an attractive contact potential, to have a bound state. 
Bound states correspond to zeroes of the inverse T-matrix~\cite{balatsky_RMP}, which takes the form $T^{-1}(\omega) \sim 1/J - \sqrt{M_x M_y M_z} (A - B \omega)$, where $A, B$ are expressions that \emph{do not} depend on the impurity mass but do in general depend on a high-energy cutoff. We eliminate this cutoff-dependence using our knowledge of the infinite-mass (i.e., pinned-impurity) bound state energy $E_\infty \simeq \Delta J^2 N(0)^2$. We then find that the threshold $J_0$ for the molecular state to exist is

\beq\label{threshold}
J_0 N(0) \simeq \frac{(2m)^{3/2}}{M } \left( \frac{1}{M} + \frac{v_F^2}{\Delta} \right)^{1/2} \sqrt{\frac{E_F}{E_\infty}}
\eeq
whereas, for $|J| > J_0$, the molecular binding energy (measured from the gap edge) is

\beq\label{ebound}
E_b \simeq - E_\infty (1 - J_0/|J|).
\eeq
Here, $m$ is the fermion mass; $N(0) \sim m k_F$ is the density of states per unit volume at the Fermi level. 
Moreover, the relative-coordinate wavefunction of the molecule decays exponentially, with a characteristic real-space size of $1/\sqrt{2 M |E_b|}$ in the directions tangent to the Fermi surface and $v_F/\sqrt{2 \Delta |E_b|}$ (generally much longer) in the normal direction.

The midgap Shiba molecule's dispersion follows analogously. Rotational invariance implies that there is a bound state of equal binding energy for every $\mathbf{P}_0$ whose magnitude is $k_F$. Thus, the molecule has a spherical dispersion minimum centered at $k_F$. The mass in the direction normal to the Fermi surface is simply the sum of the impurity mass and the inverse curvature of the quasiparticle dispersion: $
M^\perp_{mol.} = M + \Delta/v_F^2$.
%

%

\emph{Parity-changing transition}. For small $J$ the bound state energy is close to the gap edge. Thus, the midgap Shiba molecule costs energy $\sim \Delta$ relative to the atomic-branch ground state (which has no quasiparticles). As $J$ increases, the gap between atomic and molecular branches closes, and they cross at some $J_c$~\cite{salkola_PRB, sakai}. In this regime, $\Delta$ cannot be treated as large; however, we retain the assumption that the recoil $\mathcal{E}$ is a large scale (Fig.~\ref{structure}). Specifically, we assume $J, \Delta \ll \mathcal{E} \ll E_F$. We then find the molecular energy by computing the T-matrix for impurity-quasiparticle scattering in the ladder approximation (Supplemental Material). 
We find that the critical coupling obeys

%
%

\beq\label{transit}
J_c \simeq [k_F^2/(M \Delta)] J_c^\infty,
\eeq
where $J_c^\infty \sim 1/N(0)$ is the fixed-impurity transition point~\cite{balatsky_RMP}. 
The $M$-dependence follows from phase-space considerations. In the fixed-impurity limit, the bound state involves quasiparticle states from the entire Fermi surface (Fig.~\ref{structure}, left). By contrast, for a mobile impurity, recoil limits accessible quasiparticle states to a patch of transverse dimension $\sim \sqrt{M E_b}$. This phase-space reduction means the critical $J_c$ needed for a given bound-state energy is increased by a factor $(k_F/\sqrt{M E_b})^2$ relative to the fixed impurity case.



\begin{figure}[tb]
\begin{center}
\includegraphics{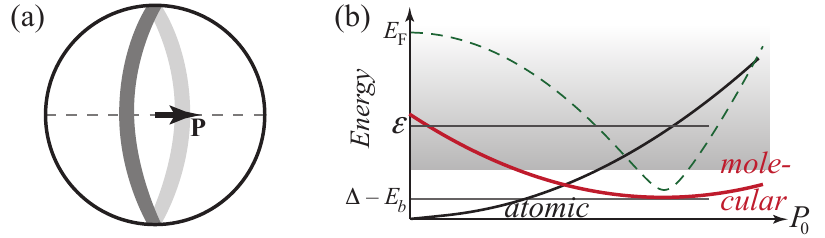}
\caption{(a)~
An impurity with momentum $\mathbf{P}$ is likeliest to create quasiparticle pairs along the shaded strip of the Fermi surface; these excitations, being tangent to the Fermi surface, are infinitely massive.
(b)~Schematic dispersions of the lowest states in the atomic (blue) and molecular (red) branches for weak coupling $J N(0) \simeq 0.2$; here, effective-mass corrections are small. For $\Delta \ll \mathcal{E}$ the branches cross. At high energies (shaded) these branches merge into the multi-particle continuum. }
\label{kinematics}
\end{center}
\end{figure}

\emph{Effective mass}. So far, we have explored the effects of the impurity recoil on the bound-state spectrum. 
We now discuss how the fermions affect the impurity recoil via polaronic effective-mass shifts~\cite{leelowpines}, involving processes in which the impurity emits and reabsorbs virtual quasiparticle pairs. Because each quasiparticle pair costs an energy $\agt 2\Delta$, the creation of many pairs is suppressed (i.e., there is no orthogonality catastrophe). Moreover, for kinematic reasons, these pairs are likeliest to lie on the dispersion minimum (Fig.~\ref{kinematics}(a)). Thus, the quasiparticle-pair energy $\approx 2\Delta$, and the perturbative impurity energy shift is

\begin{equation*}
 -J^2 (m \Delta)^{3/2} \!\! \int \! d^3 q \bigg / \left\{ \frac{\mathbf{P}^2}{2M} - \left[\frac{(\mathbf{P- q})^2}{2M} + 2 \Delta\right] \right\}.
\end{equation*}
The ultraviolet divergence in this expression can be eliminated by accounting for the high-$q$ behavior of the interaction vertex. 
For computing the effective mass, one need not regularize this divergence: the second derivative at $P = 0$ converges, yielding the effective mass

\beq\label{effmass1}
M^* \simeq M\{1 + 3/(\sqrt{2} \pi^4) J^2 \Delta (m M)^{3/2}\}.
\eeq
The effective mass of the midgap Shiba molecule (obtained similarly) is 

\beq
M_{mol.}^{\perp, *} \!\!\! \simeq \! M^{\perp}_{mol.} \!\! [1 \! + \! 1/(16\sqrt{2} \pi^5) J^2 k_F^2  (m^3 M^{\perp}_{mol.})^{1/2} ]
\eeq
For larger $J$, one must go beyond perturbation theory; as detailed in the Supplemental Material, one can find $M^*$ self-consistently by~(a)~replacing the bare interaction with the T-matrix~\cite{braaten}, and~(b)~including processes in which the impurity emits multiple quasiparticle pairs. When $J N(0) \gg 1$, we find the $J$-independent result

\beq\label{effmass2}
M^* \simeq m [E_F^2 /\{ \mathcal{E} (\Delta - E_b)\}]^{2/3}
\eeq
The \emph{dressed} impurity retains a Mexican-hat dispersion (and our calculations remain self-consistent) as long as $\mathcal{E} \agt \sqrt{\Delta E_F}$, i.e., in the BCS limit. This result, although derived specifically for the Shiba molecule, is in fact a \emph{general} result for polaron problems in which the ``bath'' dispersion is Mexican-hat-shaped (e.g., Rashba spin-orbit coupled systems~\cite{stanescu}). 

Combining results for $E_b$ and $M^*$, one can construct the full dispersion of the impurity [Fig.~\ref{kinematics}(b)]. At small $P_0$ the lowest-energy state is in the atomic branch with effective mass $M^*$; as $P_0$ increases, the atomic and molecular branches cross (provided that $\Delta \alt \mathcal{E}$), and the dispersion near $k_F$ is Mexican-hat shaped with a curvature $M_{mol.}^{\perp, *}$. 
At momenta $(2M^* \Delta)^{1/2} \alt P_0 \alt k_F - (M_{mol.}^{\perp, *} \Delta)^{1/2}$, the impurity radiates into the two-quasiparticle continuum. As $J$ increases, the minimum of the molecular branch at $P_0 = k_F$ decreases through zero, and the ground state changes via a first-order phase transition (states of different fermion parity cannot mix).  Thus the parity transition discussed here resembles polaron transitions~\cite{stojanovic, *berciu} in which the ground-state momentum changes abruptly.

\begin{figure}[tb]
\begin{center}
\includegraphics{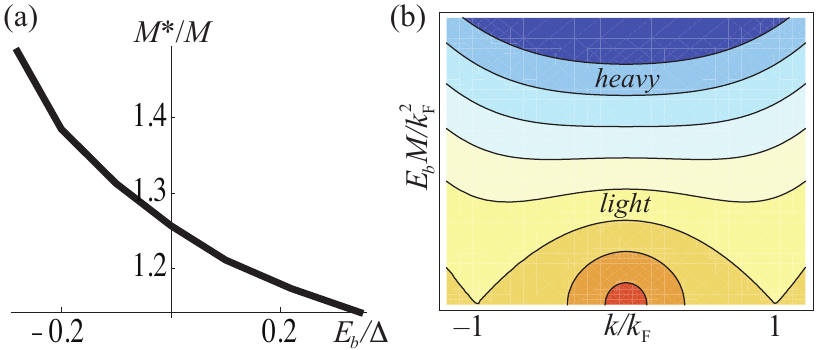}
\caption{(a)~Effective mass in one dimension as a function of Shiba state energy, computed using the T-matrix approach (see Supplemental Material); $M^*$ evolves smoothly across the parity-changing transition at $E_b = 0$. (b)~Crossover from light impurities (two dispersion minima) to heavy impurities (single minimum).}
\label{tmatrix}
\end{center}
\end{figure}

\emph{Heavy-impurity limit}. We now discuss the crossover between ``light'' and ``heavy'' impurities. 
For simplicity we work in one dimension; here, a Shiba state forms at each Fermi point. These states are mixed by an interaction matrix element $\sim E_b$; however, scattering across the Fermi surface costs $\sim \mathcal{E}$. This $2 \times 2$ Shiba subspace has the Hamiltonian

\beq
\mathbb{H} = \left(\begin{array}{cc} \frac{1}{2M}(P_0 - k_F)^2 + E_b & E_b \\ E_b & \frac{1}{2M}(P_0 + k_F)^2 + E_b \end{array} \right). \nonumber
\eeq
The smallest eigenvalues of $\mathbb{H}$ occur for $P_0 \simeq \pm k_F$ (the case discussed above) when the recoil is large; for a heavy impurity, however, the dispersion minimum moves to $P_0 = 0$ [Fig.~\ref{tmatrix}(b)]. A similar crossover occurs in any dimension. 

\emph{BEC-BCS crossover}. The analysis above assumed $E_F \gg \Delta$; this is valid in the BCS limit. However, our qualitative conclusions are based on the observation that the quasiparticle dispersion has a minimum at some nonzero momentum $k_0$. This remains true in the unitary regime but with $k_0 < k_F$; therefore, our main results (in particular, the Mexican-hat dispersion) should extend to this regime if $k_F$ is replaced with $k_0$. 
Deep in the BEC regime, the quasiparticle dispersion has a minimum at $k = 0$; our results do not apply here.


\emph{Experimental implementation}. The system discussed here can be realized in two-species atomic mixtures in which at least one species is fermionic. A promising realization involves Li-Cs mixtures~\cite{repp_heteronuclear, colin-pra}, in the magnetic-field range of 834-900 G. The Li atoms form a BCS superfluid, while the Cs-Li interaction can be tuned through various heteronuclear Feshbach resonances~\cite{chin_Li6, chin_RMP}. The impurity recoil $\mathcal{E} \approx E_F/5$, while $\Delta \sim$ 0.01-0.1 $E_F$~\cite{chin_RF}. The molecular dispersion can be \emph{directly} probed using momentum-resolved radio-frequency (rf) spectroscopy, as follows. Suppose the impurity-fermion scattering length is negative for $\uparrow$ fermions and positive for $\downarrow$ fermions. Then $\uparrow$ fermions form a Shiba state; moreover, a Feshbach molecule of the impurity and a $\downarrow$ fermion must exist. One can use an rf pulse to flip the spin state and drive transitions between the midgap Shiba molecule and the Feshbach molecule. Because the rf pulse is momentum-conserving, one can map out the dispersion relation of the midgap Shiba molecules by measuring the momentum of the Feshbach molecules (through time-of-flight imaging) as a function of frequency~\cite{jin_momentum_resolved_rf, levin_momentum_resolved_rf}. 

\emph{Outlook}. We have argued that a moving magnetic impurity in a Fermi superfluid can capture a quasiparticle and form an exotic midgap Shiba molecule with a Mexican-hat dispersion minimum; as this dispersion minimum maps out the Fermi surface, one can easily tune its shape by putting the fermions in an optical lattice. Depending on the impurity statistics, the molecule can be bosonic or fermionic. 
Moreover, we expect the intermolecular exchange interactions to be exotic. 
Qualitative aspects of these interactions can be deduced from recent work on pinned impurities~\cite{yao13, yao14}. Molecules interact by exchanging either continuum quasiparticles or Shiba states. Remarkably, for a moving impurity the Shiba-state exchange interaction is strongly \emph{angle-dependent} because the molecular wavefunctions are anisotropic, as discussed above: molecules with center of mass momenta $\mathbf{\hat{n}}, \mathbf{\hat{n}}'$ interact more strongly when $\mathbf{\hat{n}} \parallel \mathbf{\hat{n}}'$ than when $\mathbf{\hat{n}} \perp \mathbf{\hat{n}}'$. A quantitative treatment of these interactions will be given elsewhere. 
Such interactions make midgap Shiba molecules promising platforms to study the interplay between spherical dispersions and structured interactions~\cite{spielman-review, *wilson:meron, *sg:quasi}. Note that a sufficiently high density of impurities might alter the character of the superfluid, favoring a modulated gap~\cite{zapata, *arovas_kink}.
Finally, while we discussed impurities in s-wave superfluids, even more unusual properties might be realizable with impurities in unconventional (e.g., topologically paired) superfluids~\cite{huihu, *sau2013, gurarie, masatoshi, shoucheng_review}.

\emph{Acknowledgments}. We thank Cheng Chin, Gergely Zarand, Brian Skinner, and Michael Knap for helpful discussions. This work was supported by the Harvard Quantum Optics Center (S.G.), the ARO-MURI Non-equilibrium Many-body Dynamics grant (C.V.P.), the Harvard-MIT CUA, the DARPA OLE program, the AFOSR-MURI on New Quantum Phases of Matter, the ARO-MURI on Atomtronics, and the ARO MURI Quism program (E.D.).


%

\begin{widetext}

\section{Supplemental Material: Mobile magnetic impurities in a Fermi superfluid: a route to designer molecules}

\begin{center}\emph{Sarang Gopalakrishnan, Colin V. Parker, and Eugene Demler}\end{center}


In what follows, we introduce two self-consistent approaches to computing the bound state energy and the impurity effective mass. First, we discuss an approach that is tailored for light impurities in three (or two) dimensions, and yields good results in both the weak and strong coupling limits; next, we turn to the one-dimensional case and discuss an approach that is reliable across the parity-changing transition whenever the impurity is sufficiently massive. Together, these approaches support the physical argument in the main text that the parity-changing transition should not be accompanied by any divergences, as it is strongly first-order.


\section*{Higher-dimensional T-matrix approach (light impurity)}

\subsection{Estimate of the parity transition}

In this section we begin with the Hamiltonian [Eq. (1) of main text] and compute the impurity-fermion T-matrix treating the BCS coherence factors [Eq.~(2) of main text] \emph{exactly} rather than approximately. This allows us to estimate the transition between the phase in which the molecule is an excited state and that in which the molecule is the ground state.

\begin{figure}[b]
\begin{center}
\includegraphics{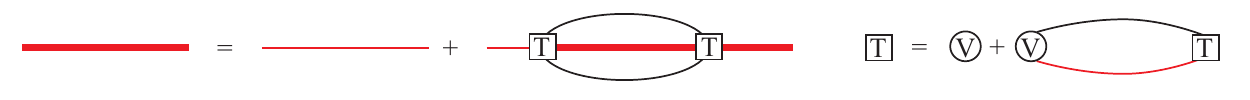}
\caption{Upper panel: equations for the self-consistent impurity propagator and for the T-matrix in the ladder approximation. Lower panel: illustration of a typical higher-order diagram of the kind resummed by our procedure.}
\label{tmatro}
\end{center}
\end{figure}

In general the T-matrix for impurity-quasiparticle scattering is given in the ladder approximation by the equation

\beq\label{ladder1}
\hat{T}(\omega) = \hat{J} + \hat{J} \left[ \int d^d k d\Omega \hat{G}(\mathbf{k}, \Omega) D(-\mathbf{k}, \omega - \Omega) \right] \hat{T}(\omega)
\eeq
where the hats denote Nambu matrix structure; matrix multiplication is implied; $\hat{J} \equiv J \mathbb{I}$ is the interaction; $\hat{G}$ is the bare fermion (i.e., $c^\dagger c$) propagator; and $D$ is the impurity propagator,

\beq\label{impprop}
\hat{G}(\omega, \mathbf{k})  = \frac{1}{\omega^2 - (\Delta^2 + v_F^2 k^2)} \left(\begin{array}{cc} \omega - v_F (q - k_F) & \Delta \\ \Delta & \omega + v_F (q - k_F) \end{array} \right); \quad D(\omega, \mathbf{k}) = \frac{1}{\omega - k^2/(2M)}.
\eeq
To find the parity transition we need to find the poles in $\hat{T}(0)$. Once $\Omega$ is integrated over we have

\beq\label{suppT}
\int dq \, d^{d-1} k \, \frac{1}{\sqrt{\Delta^2 + v_F^2 q^2} + k^2/(2M)} \left(\begin{array}{cc} \sqrt{\Delta^2 + v_F^2 q^2} - v_F (q - k_F) & \Delta \\ \Delta & \sqrt{\Delta^2 + v_F^2 q^2} + v_F (q - k_F) \end{array} \right).
\eeq
where $q$ is the direction normal to the Fermi surface and $k$ denotes all other directions. In this expression we have assumed that $v_F^2/\Delta \gg 1/M$, i.e., we neglect the change in curvature of the Bogoliubov dispersion minimum. This inequality is equivalent to requiring $E_F/\Delta \gg m/M$, which is always satisfied in the BCS limit. 
We extract the $M$-dependence from this integral by substituting the variable $s = k/\sqrt{M}$. One sees then that the integral is given by $M^{(d-1)/2} \times F(\Delta, v_F)$, where the latter function depends exclusively on fermionic parameters and not on $M$. The remaining integral is formally divergent, but this divergence can be eliminated, and $F$ can be determined, if one requires the bound state energy to go smoothly to its infinite-mass limit. This yields the result in the main text.

%

\subsection{Effective mass at strong coupling}

We now turn to the effective mass $M^*$ of the impurity at strong coupling. We make two assumptions: (i)~that the superfluid is in the BCS limit, so that $E_F$ is much greater than any other energy scale in the problem; (ii)~that the impurity is sufficiently light that its renormalized recoil energy $\mathcal{E}^* \equiv 2 k_F^2/M^*$ remains greater than the coupling scale. Condition (ii) will be checked for self-consistency at the end of the calculation. When conditions (i) and (ii) are satisfied, the effective mass can be computed for arbitrary coupling if one replaces the bare interaction vertex with the T-matrix and resums the diagrams shown in Fig.~\ref{tmatro}.
These diagrams dominate because, in the light-impurity and BCS limits, the impurity can only emit and absorb pairs of quasiparticles with nearly opposite momenta. The diagrams in Fig.~\ref{tmatro}, in which successive each quasiparticle pair can lie anywhere on the dispersion minimum, are parametrically more important (by a factor $\sim \Delta/\mathcal{E}$) than diagrams in which the quasiparticle lines cross. 

Specifically, the effective mass is given by the equation:

\bea\label{ladder2}
\frac{1}{M^*(\hat{T})} & = & \frac{1}{M} - \frac{d^2}{dP^2} \int d\omega\, d\Omega \, d^3 k \, d^3 q\,  D_{M^*}(-\Omega - \omega, \mathbf{P} - \mathbf{q} - \mathbf{k}) \nonumber \\ & & \quad \qquad \times \mathrm{Tr} \left[ \hat{G}(\omega, \mathbf{k})\hat{T}(\omega) \hat{G}(\Omega, \mathbf{q})  \hat{T}(\omega) \right].
\eea
where the impurity propagator on the right-hand side is to be computed using $M^*$. The equations~\eqref{ladder1}, \eqref{ladder2} together define a self-consistent theory incorporating the effects of the impurity recoil and the quasiparticle rearrangement on each other. 

In this self-consistent approach, there are two ways for the impurity to emit a quasiparticle: either it can virtually occupy a Shiba state and a continuum state, or it can virtually occupy two continuum states. (There cannot be two Shiba states at once as at least one of the quasiparticles must have the wrong spin to form a Shiba state.) We first briefly consider processes in which an impurity creates a Shiba state and a continuum state. It is straightforward to see that, even in the extreme limit where the zero-quasiparticle state hybridizes resonantly with such a state \emph{and} the molecule is infinitely massive, this channel at most increases the effective mass by a factor of two. (This is the standard result for the hybridization between a dispersing band and a flat band.) As we shall see, the continuum channel is parametrically more important at strong coupling.

We now discuss the behavior of this continuum channel. From the discussion of the pair-creation diagram in the main text, we know that the dominant quasiparticle frequencies contributing to Eq.~\eqref{ladder2} are $\alt \Delta$. Therefore, we replace the T-matrix in this equation with its value at $\Delta$, i.e., the bottom of the two-particle continuum; this can be expressed in terms of an impurity-quasiparticle scattering length $a$~\cite{braaten}

\beq
T(k) \simeq \frac{1/M^*}{a^{-1} + i k}
\eeq
Crucially, in the regime of interest (i.e., that of strong coupling, where a midgap Shiba state is present), the scattering length is given by $a \sim 1/\sqrt{2 M^* (\Delta - E_b)}$. Thus, in terms of the bound state energy, the T-matrix element for pair creation is then given in the regime of interest by

\beq
T(\Delta) \simeq \frac{1}{2 \sqrt{\Delta - E_b} M^* \sqrt{\Delta/v_F^2}},
\eeq
where we have substituted in the anisotropic mass from the main text. We emphasize that this relation does \emph{not} rely on the ladder approximation for the T-matrix. Now, we evaluate Eq.~\eqref{ladder2}, to arrive at the expression

\beq
\frac{1}{M^*}  = \frac{1}{M} - \frac{3}{\sqrt{2} \pi^4} T(\Delta)^2 m^{3/2} (M^*)^{1/2} \Delta \simeq \frac{1}{M} - \alpha \left(\frac{m}{M^*}\right)^{3/2} \frac{v_F^2}{(\Delta - E_b)},
\eeq
where $\alpha$ is a numerical constant of order unity. When $E_F$ is sufficiently large compared with the other scales, one finds that 

\beq
M^* \simeq m (M v_F^2/(\Delta - E_b))^{2/3} \sim m \left(\frac{E_F^2}{\mathcal{E} \Delta}\right)^{2/3}
\eeq
For the theory to remain self-consistent, we require that $\Delta \ll k_F^2/(2M^*)$ (otherwise it would not be legitimate to use the light-impurity limit for the T-matrix calculation). One can easily see that this self-consistency condition is satisfied whenever $\mathcal{E}^2 \gg \sqrt{\Delta E_F}$, i.e., for relatively light impurities in the BCS limit. 

This self-consistent approach relies on the physically reasonable assumption that successive pair-creation events are uncorrelated. This assumption is also made by various intermediate-coupling theories such as Ref.~\cite{leelowpines}; in contrast with such theories, however, we also include the ``vertex corrections'' (i.e., the T-matrix ladder diagrams) that are necessary to account correctly for the existence of the bound state. These corrections are much more important here than in the standard polaron problem because the impurity-fermion interaction is \emph{quadratic} rather than linear in the fermions.

\section*{One-dimensional T-matrix approach (heavy impurity)}

We now consider the case of a one-dimensional system. In this case, the dispersion minimum consists of two points rather than a surface. The resulting qualitative differences are: (i)~a midgap Shiba molecule exists for arbitrarily weak coupling, and has a binding energy $E_b \simeq - J^2 / (1/M + v_F^2/\Delta)$; (ii)~the perturbative effective-mass correction goes as $J^2/M$, and \emph{decreases} as the bare mass is increased [in contrast with the three-dimensional result Eqs.~\eqref{effmass1}, \eqref{effmass2}]. In one dimension, we can explicitly calculate (using a self-consistent method described below) the evolution of the bound-state energy and effective mass; as shown in Fig.~4 of the main text, the effective mass evolves smoothly across the parity-changing transition. 

We now introduce the self-consistent T-matrix procedure. In effect, this method treats $J$ exactly and involves resummed perturbation theory in $1/M$, the inverse impurity mass. We begin by performing a standard, exact polaron transformation~\cite{leelowpines} on $\mathcal{H}$; it then takes the form
\beq
\mathcal{H}' = H_{\mathrm{BCS}} + \sum_{kk' \sigma} \frac{(V + J \sigma)}{\mathcal{V}} c^\dagger_{k} c_{k'} + \mathrm{h.c.}  + \frac{1}{2M} \left(P_0 - \sum_{k \sigma} k c^\dagger_{k \sigma} c_{k \sigma} \right)^2
\eeq
In this equation, the first line is quadratic in the fermion operators, and corresponds to the solvable limit of a fixed ($M = \infty$) impurity. The second line includes a term that is \emph{quartic} in the fermions; this quartic term renormalizes both the dispersion and the scattering of the fermions, via the diagrams in Fig.~\ref{diagrams}. In general, the renormalized scattering will be $k$-dependent; however, we note that $k_F$ is much larger than the other momentum scales involved, which justifies approximating the total fermion momentum as the difference between the number of right- and left-movers, i.e.,

\beq\label{Lmovers}
\sum_{k \sigma} k c^\dagger_{k \sigma} c_{k \sigma} \rightarrow  k_F \sum_{k, \sigma, \tau = \pm 1}  \tau c^\dagger_{k \sigma \tau} c_{k \sigma \tau}.
\eeq
In this approximation, the T-matrix has a $2\times 2$ matrix structure in $(L,R)$ space but no other momentum-dependence. 

We are now equipped to write out a self-consistent set of equations for the fermion Green's function and the T-matrix, shown diagrammatically in Fig.~\ref{diagrams}. As we would like to discuss the even-sector effective mass even when the even sector is not the ground state, we shall work within the Keldysh framework; thus, all the Green's functions in the procedure are the retarded component, except for the Green's function in the loop, which is the Keldysh component. (The Keldysh component of the Green's function is essentially a product of the spectral function and the distribution function; we choose the distribution function to be the ground-state Fermi function \emph{except} that we require the Shiba state to remain unfilled regardless of its energy.) 

\begin{figure}[htbp]
\begin{center}
\includegraphics{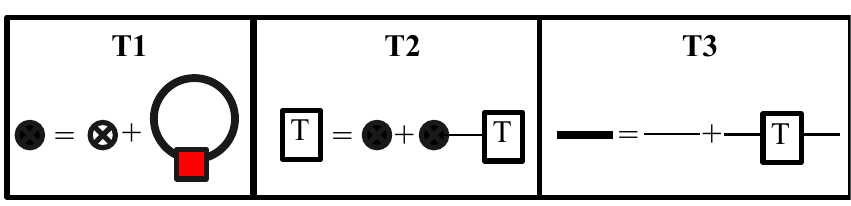}
\caption{Diagrammatic overview of the self-consistent T-matrix procedure. The circled X denotes the interaction $J$; the red square, the interaction $\sim 1/2M$; the thin lines are bare fermionic propagators; the bold lines are full fermionic propagators; and the other symbols are defined through the system of equations T1-T3. T1 shows how the impurity-fermion interaction vertex is renormalized, to lowest order in $1/M$, by impurity recoil. T2 accounts for repeated scattering events using this renormalized vertex by promoting it into a T-matrix, $T$. T3 constructs the full Green's function from the T-matrix via the standard relation~\cite{braaten} $G = G_0 + G_0 T G_0$.}
\label{diagrams}
\end{center}
\end{figure}
From these equations, it is straightforward to find the binding energy of the Shiba state (as one simply looks at the poles of the T-matrix in Eq. T2). 
In addition, one can extract the effective mass from this diagrammatic system, using the following exact relations: 

\beq
\frac{d E}{dP} = v = P/M^*; \quad \frac{d E}{dP} = \frac{1}{M} \left\langle P - \sum_k k c^\dagger_k c_k \right\rangle.
\eeq
Physically, this equation states that the group velocity of the polaron (viz. $P/M^*$) is on average the same as the velocity of the impurity, which is its momentum divided by its bare mass. 
Once the impurity Green's function is computed from diagram T3 in Fig.~\ref{diagrams}, the expectation values on the right-hand side of this equation are known, and therefore it is straightforward to compute the effective mass. 

\begin{figure}[htbp]
\begin{center}
\includegraphics[width = 0.45\textwidth]{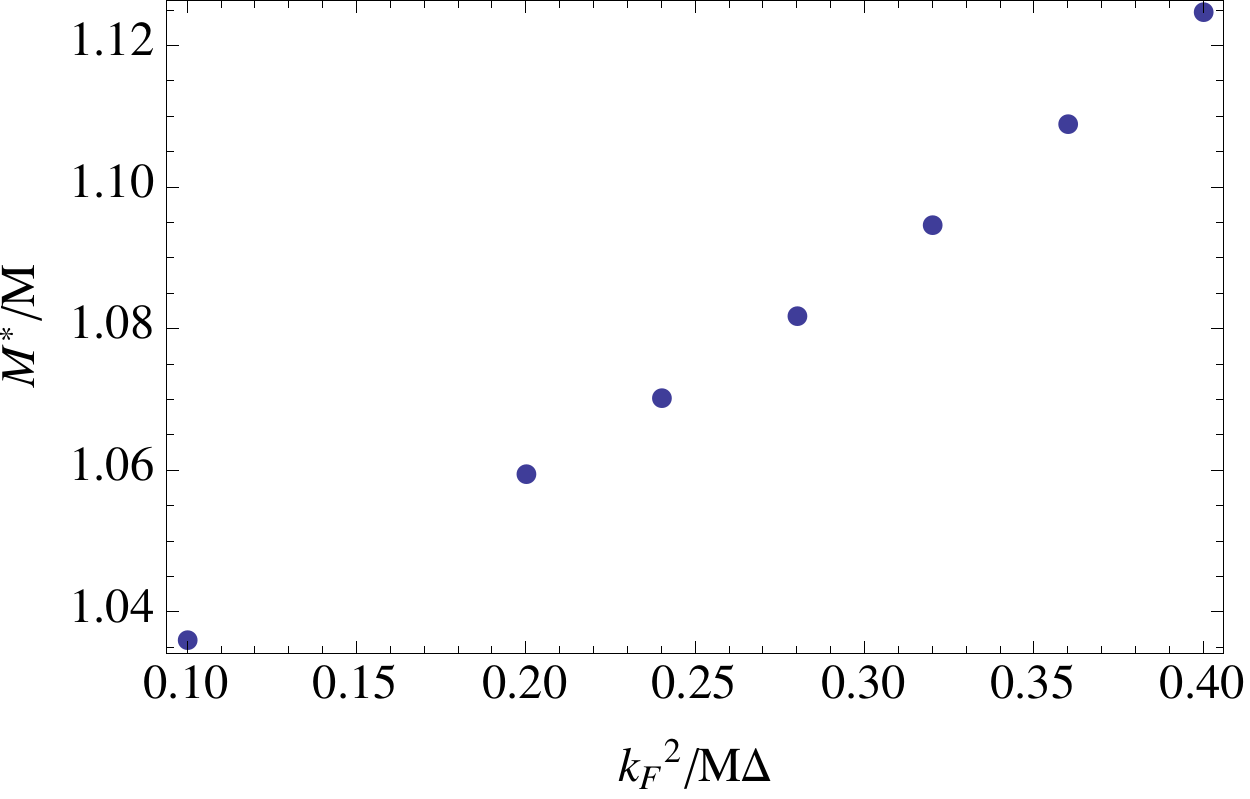}
\includegraphics[width = 0.45\textwidth]{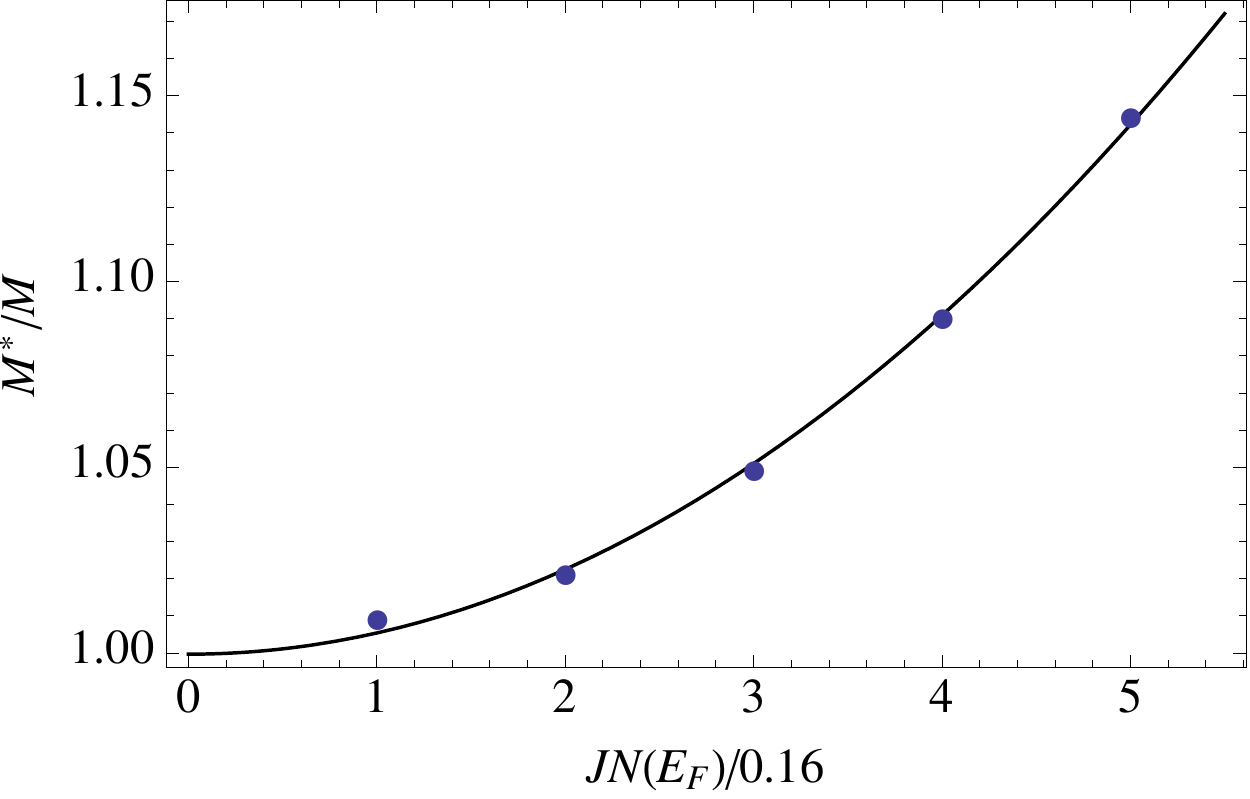}
\caption{Dependence of the effective mass correction on the bare mass and on the coupling; there are small but noticeable deviations from the $J^2/M$-dependence predicted by perturbation theory.}
\label{massdep}
\end{center}
\end{figure}
Figs.~\ref{massdep} show the typical dependence of the effective-mass correction on the bare mass and on $J$. These results are remarkably close to those obtained via perturbation theory (as in the main text):

\beq\label{1dpert}
\frac{M^*}{M} - 1 \simeq \frac{J^2 \Delta}{M v_F^4} \nonumber
\eeq

\subsection*{Generalization of heavy-impurity approach to higher dimensions}

In principle the self-consistent approach discussed here can be generalized directly to higher dimensions. For simplicity we consider the case of two dimensions. The key step is to replace Eq.~\eqref{Lmovers} with the substitution

\beq
\sum_{\mathbf{k}} \mathbf{k} c^\dagger_{\mathbf{k}} c_{\mathbf{k}} \rightarrow k_F \sum_{k \theta} \cos(\theta) c^\dagger_{k\theta} c_{k\theta}.
\eeq
One can then proceed as above, except that the renormalized interaction becomes a continuous function of $\theta$ instead of just acquiring a $2 \times 2$ matrix structure. Note that this approximation also cures the ultraviolet divergences that would arise if one tried to work with the full theory. 

We have not followed this route further in the present work because our analysis of the (computationally simpler) one-dimensional system suggests that the convergence of this approach is good only for $k_F^2/(M\Delta) \alt 1$, which is the opposite regime to that considered in the main text. However, we hope to adapt this approach to study the crossover between heavy and light impurities in future work.

\end{widetext}

\end{document}